# Modeling an optical magnetometer with electronic circuits – Analysis and optimization

**Przemysław Włodarczyk[a], Szymon Pustelny[b,c], Jerzy Zachorowski[b], and Marcin Lipiński[a]**

[a] *Department of Electronics, AGH University of Science and Technology*
Mickiewicza 30, 30-059 Krakow, Poland

[b] *Institute of Physics, Jagiellonian University*
Reymonta 4, 30-059 Krakow, Poland
E-mail: pustelny@uj.edu.pl

[c] *Department of Physics, University of California at Berkeley*
Berkeley, California 94720-7300, USA

ABSTRACT: Optical magnetometers are currently able to achieve magnetometric sensitivities below 1 fT/Hz$^{1/2}$. Although such sensitivities are typically obtained for ultra-low-field measurements, a group of optical magnetometers allows the detection of the fields in much broader dynamic range without a significant compromise in the sensitivity. A particular example of such a device is the magnetometer exploiting amplitude-modulated nonlinear magneto-optical rotation. It enables to measure a magnetic field via detection of a polarization state of light traversing a medium subjected to the field. In this paper, an electronic-circuit analogue of such the magnetometer is elaborated. Its operation is investigated with an electronic-circuit design software, which allows to study the "magnetometer" performance as a function of various parameters. The ability to automate operation of the magnetometer and automatically track "magnetic field" is demonstrated. The simulations are verified with experimental results obtained with the magnetometer operating in one of the investigated arrangements.

KEYWORDS: Numerical simulations of electronic circuits; Magnetometry; Magnetic field, Magneto-optical rotation, SPICE

## Contents

## 1. Introduction

Magnetic-field measurements are at the heart of numerous modern scientific and commercial applications. For example, such measurements are employed in research in physics, chemistry, and geology, but also in electronics, surveys for natural resources, and medicine. Each of the applications imposes different requirements on a magnetic-field sensor therefore modern magnetometric techniques exploit different physical phenomena are used. The techniques offer distinct sensitivities, dynamic ranges, and bandwidths of the sensors, as well as their different sizes, prices, and maintenance costs [1].

    Currently, the most sensitive magnetic-field measurements are provided by optical magnetometers. Such devices utilize a magnetic-field dependent change of a specific property of light propagating through a medium subjected to an external magnetic field (see Ref. [2] and references therein). The best of the magnetometers reach a sensitivity of 160 aT/Hz$^{1/2}$ in a dynamic range of 100 nT [3]. Such sensitivity and dynamic range require compensation of external, uncontrolled magnetic fields. Nonetheless, a group of optical magnetometers allows for measurement of stronger fields. In particular, the devices may reach important for many applications fields comparable to the Earth magnetic field (~50 µT), which allows for operation of the devices in a magnetically unshielded environment. A specific example of a magnetometer enabling higher-field measurements is the device exploiting amplitude-modulated nonlinear magneto-optical rotation (AMOR) for detection of the field [4]. For example, the operation of the AMOR magnetometer in a dynamic range of ~100 µT with a sensitivity of $4.3\times10^{-13}$ T/Hz$^{1/2}$ was demonstrated in Ref. [5].

The AMOR magnetometer may work in a few different experimental arrangements. The arrangements are investigated in this paper by elaborating their electronic-circuit models. The models are investigated using an electronic-circuit simulation software - Simulation Program with Integrated Circuit Emphasis (SPICE). It allows us to study diverse aspect of the magnetometer operation and optimize the device with respect of its automated "magnetic-field" tracking. Based on the simulations, a real device with such an automatic tracking of magnetic fields is set up and its performance is investigated.

## 2. Optical magnetometer based on amplitude-modulated nonlinear magneto-optical rotation

### 2.1 Real system

Resonant interaction of intense, linearly polarized light with a medium may lead to generation of an optical anisotropy in the latter [6]. In particular, the anisotropy may manifest as a rotation of a polarization plane of the light when the angle between the incident-light polarization and a medium anisotropy axis is neither 0° nor 90° (circular birefringence) [7]. In magneto-optically active media, the angle is introduced by application of an external magnetic field as the anisotropy axis rotates around the field (Fig. 1). In the case of unmodulated light, a strong, static, nonzero rotation signal is observed for weak magnetic fields only [8]. However, when the light is amplitude modulated the signal may also be observed at stronger fields [9]. In such the situation, the polarization orientation is periodically modulated due to continuous rotation of the anisotropy axis around the field, and its frequency is determined by the magnitude of the magnetic field $B$ [7]. The polarization-rotation signal is the largest when linearly polarized light is modulated at twice the precession frequency

$$\Omega_m = 2\Omega_L = kB, \quad (1)$$

where $\Omega_m$ is the modulation frequency, $\Omega_L$ is the magnetic-precession frequency (the Larmor frequency) and $k$ is the material constant (for frequently used in optical magnetometers vapor of $^{87}$Rb takes the value 14 kHz/μT). Equation (1) allows for magnetic-field measurements in a broad dynamic range and is facilitates in the, so-called, AMOR magnetometry.

Block diagram of a typical AMOR magnetometer is shown in Fig. 2. The figure presents two modes of magnetometer operation: (a) passive mode, where light modulation is induced by an external oscillator [5], and (b) self-oscillating mode [11], where an output signal of the magnetometer is used for light modulation. While the first mode yields better signal-to-noise ratio due to suppression of noise at undesired frequencies using lock-in detection, the second mode allows automation of the operation and fast field tracking.

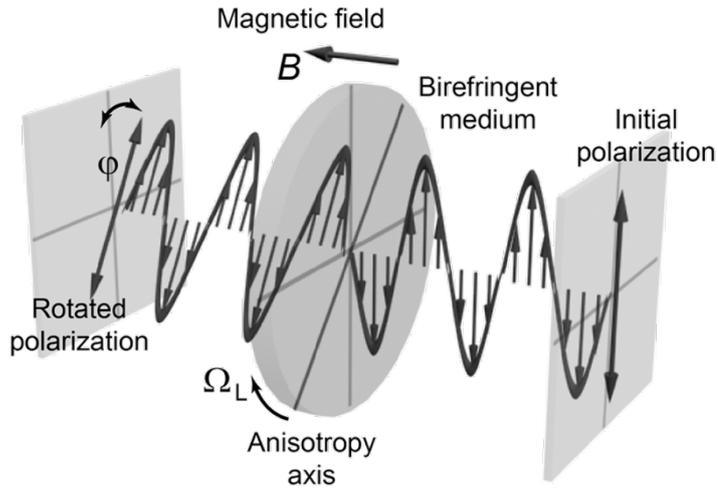

**Figure 1**. Rotation of a polarization plane of linearly polarized light in a magneto-optically active medium revealing circular birefringence. The anisotropy axis rotates around a magnetic field giving rise to static, for CW light and low field, or dynamic, for modulated light and stronger field, polarization rotation.

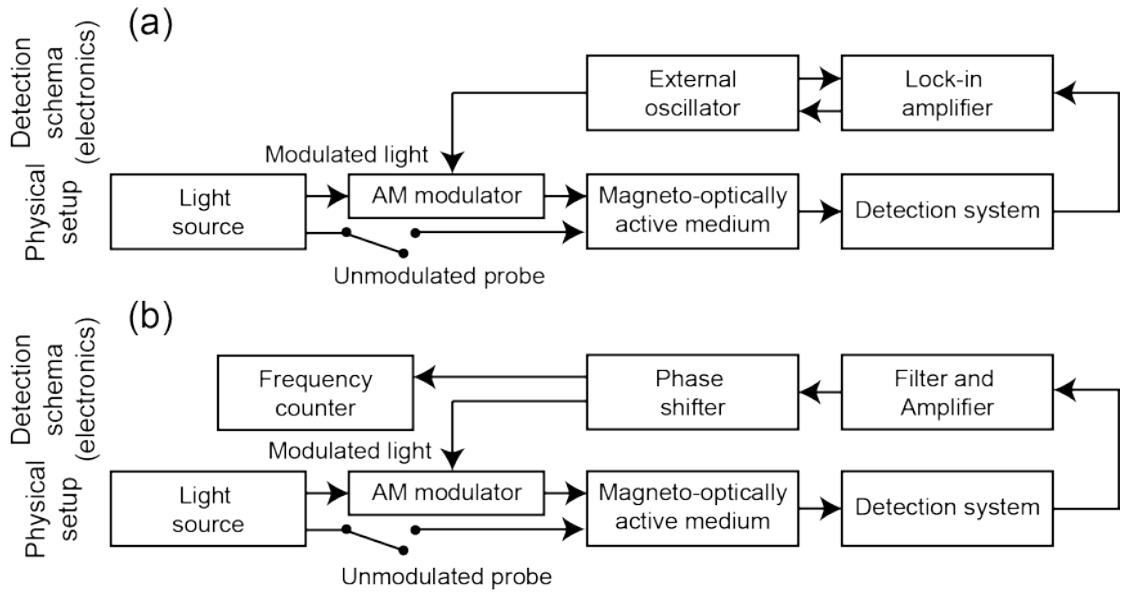

**Figure 2**. Block diagram of AMOR magnetometer in (a) passive and (b) self-oscillating modes. In both modes, the lower path extending from Light source to Magneto-optically medium corresponds to the pump-probe arrangement (see text).

AMOR magnetometers may be operated in various experimental arrangements. For instance, the arrangements may be classified regarding a number of light beams applied for generation and detection of optical anisotropy of a medium. In the simplest experimental arrangement a single, amplitude-modulated light beam is used for both generation and detection of the anisotropy (Unmodulated-probe path is open in Fig. 2). In the case, however, the time-dependent polarization rotation φ is modified by a light-modulation pattern. When an additional unmodulated light beam is used for probing the medium (closed Unmodulated-probe path – the pump-probe arrangement), the polarization-rotation angle φ is exclusively determined by evolution of the anisotropy in the external field. The magnetometers may also be classified by polarization-rotation detection schemes. In general, in all cases the detection is performed with a polarizer placed after the medium. If the polarizer axis is rotated by 45° with respect to the incident-light polarization, the difference of intensities of light beams directed into two polarizer channels provide the rotation angle φ (Appendix). If, however, the polarizer is rotated by almost 90° with respect to initial light polarization, the forward transmitted light is a second order polynomial of polarization rotation angle (Appendix).

In this paper, four AMOR-magnetometer arrangements are considered. They are the permutations of the schemes discussed above. Table 1 provides characteristics of optical signals detected in either of the arrangements (for more details see Appendix). First two rows correspond to the pump-prove arrangements, where unmodulated probe light is used. The other two rows correspond to the signals detected for the single-beam arrangements. The first and third rows are associated with a signal recorded in the balanced-polarimeter scheme, while the second and fourth with a nearly-crossed polarizer. The last column in Table 1 shows the corresponding characteristics of the electrical signal used in our simulations.

**Table 1**. Time characteristics of AMOR-magnetometer optical signals recorded in four different arrangements investigated in this paper. It was assumed that in case of probe-light modulation, the light is sinusoidally modulated with 100%-modulation depth. The last column shows electric signals used in simulations of a particular magnetometer arrangement. $V_{atoms}$ is the electric response of the RLC circuit to the modulation, which corresponds in the magnetometer to the response of a medium to light modulation and $V_{laser}$ is the signal associated with light modulation at the probing stage (note that in first and second row $V_{laser}=1$). $A$ is the amplitude of the polarization rotation, $\Theta$ is a small angle by which the polarizer is tilted from 90°, and $I_0$ is the probe[1]-beam intensity.

| Beams arrangement | Detection arrangement | Mathematical formula describing recorded signal | Function characterizing the source $B_1$ in Fig. 3 |
|---|---|---|---|
| Pump-probe Arrangement | Balanced polarimeter | $2I_0 A \sin\Omega_m t$ | $V_{atoms}$ |
| Pump-probe Arrangement | Nearly-crossed analyzer | $2I_0(A\sin\Omega_m t + \Theta)^2$ | $(V_{atoms} + \Theta)^2$ |
| Single-beam Arrangement | Balanced polarimeter | $I_0/2(1 + \sin\Omega_m t) A \sin\Omega_m t$ | $V_{atoms} V_{laser}$ |
| Single-beam Arrangement | Nearly-crossed analyzer | $I_0/2(1 + \sin\Omega_m t)(A \sin\Omega_m t + \Theta)^2$ | $(V_{atoms} + \Theta)^2 V_{laser}$ |

## 2.2 Electronic equivalent

By its nature, AMOR remains a driven damped harmonic oscillator, where the oscillation period is determined by a magnetic field. Thus a non-electronic part of an AMOR magnetometer has a simple electronic equivalent, an RLC circuit, which may be easily simulated with electronic circuit simulation software such as Spice.

---

[1] Note that in the single-beam arrangement probe light is modulated.

The left-hand circuit of Fig. 3 shows the circuit used to model response of a real medium to amplitude-modulated light. In the electronic case, the properties of the light, i.e., its intensity characteristics and frequency, are determined by the voltage $V_{laser}$. The response of the medium to the light is given by the voltage $V_{atoms}$.

In order to reproduce the AMOR signal measured in rubidium vapor in the Earth magnetic (~50 µT), the resonance frequency of the circuit is set to 700 kHz by appropriate choice of the values of the electronic elements. The width of the resonance (~30 Hz) and hence the Q factor of the circuit is adjusted by the resistivity of the resistor R.

The second part of Fig. 3 presents an electronic circuit modeling the detection system in an AMOR setup. It consists of the behavioral source $B_1$ providing signal $V_{det}=f(V_{atoms},V_{probe})$, which is determined by response of the RLC circuit (medium) $V_{atoms}$ and time characteristic of the detecting signal $V_{probe}$ (probe-light intensity). The function takes different forms depending on a particular configuration of the experimental setup (see Table 1).

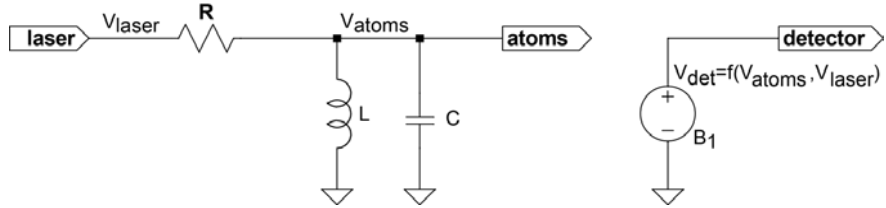

**Figure 3**. Electronic equivalent of the AMOR magnetometer with left-hand side circuit corresponding to the optically driven medium and right-hand side the detection scheme. R, L, C correspond, respectively, to the resistor, inductor, capacitor, and $B_1$ denotes the behavioral source providing the voltage $V_{det}$.

## 3. Simulations

### 3.1 General information

The electronic circuits are simulated in the time domain with LTspice IV, which is a high performance SPICE simulator, schematic capture, and waveform viewer developed by Linear Technology. In each of the considered cases, the maximum simulation time-step size is 3 ns. Such the small step is chosen to minimize the error of the transient analysis, which becomes significant when considering resonant circuits with high Q factor.

An important purpose of our analysis is to investigate the possibility of operating an AMOR magnetometer in the self-oscillating mode, which automates magnetic-field tracking. In the case, absolute values of the detected signals are not important, as they may be modified by adjusting gain in the detection system, so that we have assumed, for simplicity, that the amplitude of the oscillation of the polarization rotation $A$ is equal to 1. The angle $\Theta$ is in general a parameter, which can be freely chosen in the experimental setup. We have performed the simulations for a broad range of $\Theta/A$ ratios (between 0.001 and 1000), which modifies amplitudes of specific harmonics in the signal (see Appendix). It was found that in all the cases just an adjustment of the amplification in the system is required to achieve self oscillations. Therefore, herein the results for the case of $\Theta = A = 1$ are presented.

## 3.2 Model verification

In order to validate the circuits used for modeling of the AMOR experiment, a typical AMOR signal obtained in the pump-probe arrangement with the balanced polarimeter (top row of Table 1) are compared with the results of numerical simulations in the electronic-equivalent circuits shown in Fig. 4. Since the dependence of the polarization rotation on the modulation frequency is most accurately reproduced in the passive mode [Fig. 2(a)], a voltage control generator is used in the simulations to drive the RLC circuit. The behavior of the circuit is simulated by sweeping the generator frequency by ramping the generator input voltage $V_{ramp}$. The components $B_2$, $R_2$, $C_2$, and $B_3$, $R_3$, $C_3$ form synchronous-detection circuits and correspond to a lock-in amplifier used in the experiment. The voltages $V_{lockin}^{sin}$ and $V_{lockin}^{cos}$ correspond to the in-phase and quadrature components of the signal measured in the experiment. The capacitors $C_2$ and $C_3$ determine integration time constants and their values are set in order to obtain the best tradeoff between accuracy and simulation time. The 1-V DC voltage source between the generator and the resonant circuit assures that voltage $V_{laser}$ corresponding to light intensity does not become negative.

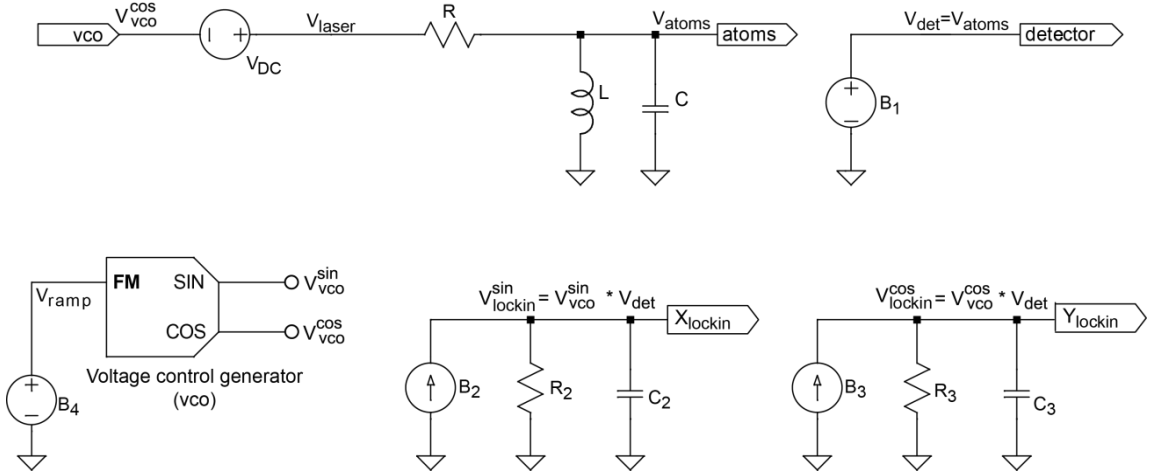

**Figure 4**. Circuit diagram and simulation parameters of the electronic equivalent of the AMOR setup with balanced polarimeter in pump-probe arrangement (upper circuits) where optical signal is detected using lock-in amplifier (lower circuits).

Figure 5(a) shows the results of the simulation versus the modulation frequency. The amplitude $R_{lockin}$ is calculated from the voltages $V_{lockin}^{sin}$ and $V_{lockin}^{cos}$ with the formula $R_{lockin} = \sqrt{(V_{lockin}^{sin})^2 + (V_{lockin}^{cos})^2}$. The signal reveals a Lorentzian shape which, for a given set of parameters, centers at ~700 kHz. The results are shown aside the experimental signal measured with the AMOR magnetometer in the pump-probe with the balanced polarimeter [Fig. 5(b)]. The signal was obtained by scanning the light modulation frequency and recording the amplitude of AMOR (for more details see Ref. [4]). The comparison shows that there is nearly perfect agreement between experiment and simulations. A small discrepancy between positions of the resonances is an artifact originating from simulations. In the transient analysis resonant-circuit response depends on the simulation step. In practice, the simulated resonant frequency of the system is lower than the theoretical value. This problem is particularly important for circuits with a high Q factor such as considered in our system.

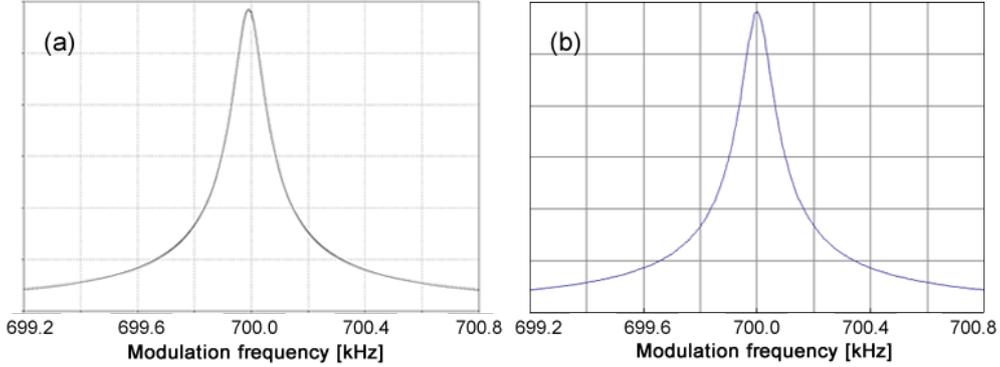

**Figure 5**. (a) Simulated electric signal generated with LTspice in the RLC circuit corresponding the pump-probe arrangement with the balanced polarimeter and (b) amplitude of the AMOR signal as a function of the modulation frequency for a longitudinal magnetic field of 50 µT. As shown in Table 1, the arrangement most accurately corresponds to the actual rotation of the polarization plane generated in a magneto-optically active medium. The simulations are performed for following values of electronic components shown in Fig. 4: $R$=200 kΩ, $R_2$=$R_3$=1 Ω, $L$=5.17 µH, $C$=10 nF, $C_2$=$C_3$=10 mF. The experimental signal was measured for the laser tuned to the F=1→F'=1 transition of $^{87}$Rb contained in anti-relaxation coated, buffer-gas-free glass cell (for more details see Ref. [11]).

### 3.3 Different detection schemes in self-oscillating mode

#### 3.3.1 Pump-probe arrangement with balanced polarimeter

Successful verification of the electronic model of the AMOR experiment allows us to proceed to investigation of the AMOR magnetometer in the self-oscillating mode. First, the simulations of the magnetometer in the pump-probe arrangement with the balanced polarimeter are performed. The arrangement is similar to that discussed in Sec. 3.2. One difference consists in replacement of the generator driving the RLC circuit with an appropriately filtered and amplified experimental signal provided by the source $B_1$. Additionally, the current arrangement does not exploit phase-sensitive detection.

Figure 6 shows the electronic circuits that are used to simulate the arrangement. A small amount of white noise $V_{noise}$ is introduced into the source $B_1$. In a real magnetometer such the noise is of fundamental origins and it arises, e.g., from the Heisenberg uncertainty principle. From the practical point of view, the noise is required to initiate the oscillations as the system promotes a particular frequency component in the signal, which eventually leads to self oscillations. In order to limit the amplitude of the driving voltage $V_{B2}$, which corresponds to an experimental limit of finite energy accessible in the system, an output signal is processed with an automatic gain control (agc) system. The agc system is formed with a peak detector consisting of the diode $D_1$, the resistors $R_1$ and $R_2$, the capacitor $C_1$ and the voltage source $B_2$. The peak detector provides the voltage $V_{agc}$ that controls the amplification $k$ of the source $B_2$, so that the amplitude of the modulating signal $V_{laser}$ remains constant, independently on the amplitude of the output signal $V_{det}$. The additional voltage source $B_3$ introduce into the system controls the switch $S_1$, whose task is to detune the resonance frequency of the RLC circuit by attaching an additional capacitance $C'$ at a given time $t_0$.

The simulation results are shown in Fig. 7. The system starts to oscillate by itself reaching the maximum oscillations within ~5 ms [Fig. 7(a)]. Since the oscillation frequency is determined by the RLC circuit, when the switch $S_1$ connects the capacitor $C'=C/2$ at $t_0 = 8$ ms, the resonance frequency of the system changes. The modification leads to instant adjustment of the oscillation frequency [see inset to Fig. 7(a)]. It demonstrates the ability to track magnetic-field changes in the arrangement.

According to theoretical considerations, the system should oscillate sinusoidally (compare to Table 1). This prediction is verified by the fast Fourier transform (FFT) of the signal [Fig. 7(b)]. As seen, there are two frequencies in the spectrum of the signal that corresponding to oscillation at $t < 8$ and $t > 8$ ms.

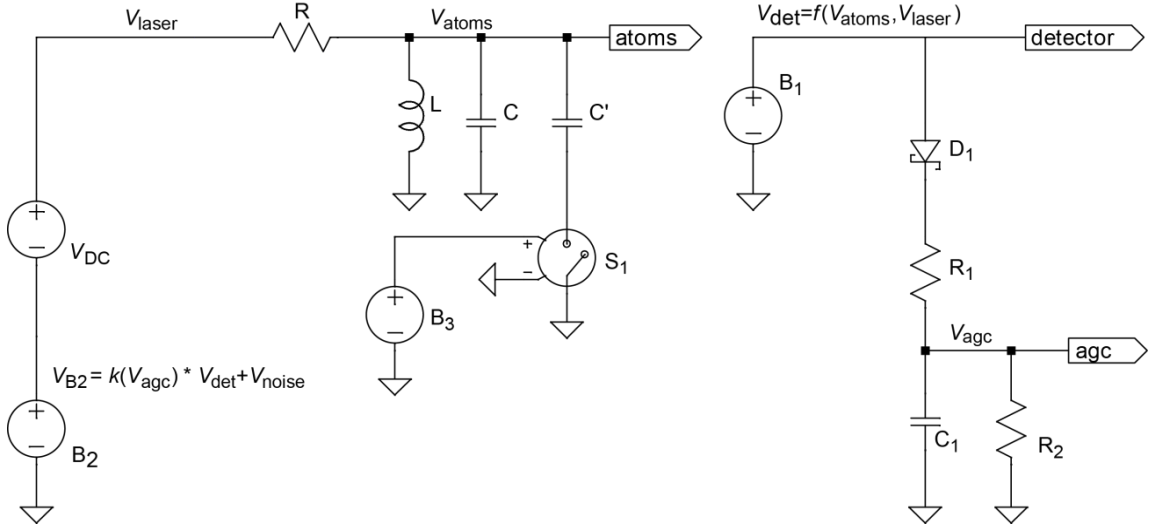

**Figure 6**. Electronic model of the pump-probe arrangement with the balanced polarimeter. $V_{det}$ is the output signal, which for first three arrangements is given by the formulas provided in Table 1.

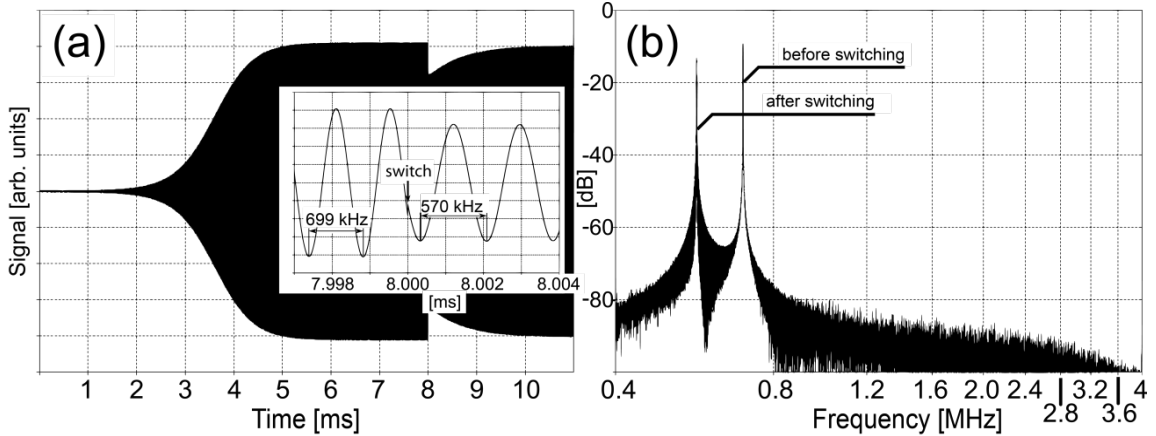

**Figure 7**. (a) Time-dependent electric signal observed in the system corresponding to the pump-probe arrangement with the balanced polarimeter. Inset shows a moment of rapid change of the RLC-circuit resonance frequency. The simulations show instant response of the system to the change. (b) FFT

of the signal with two distinct frequencies corresponding to oscillations of the system before and after 8$^{th}$ ms. The simulations were performed for following parameters: $R$=200 kΩ, , $R_1$=1 kΩ, $R_2$=100 kΩ, $L$=5,17 µH, $C$=10 nF, $C'$=$C/2$=5 nF, $C_1$=3 nF, $D_1$ is the Schottky diode type BAT54, and the rms amplitude of the noise $V_{\text{noise}} = 0.01$.

### 3.3.2 Pump-probe arrangement with nearly-crossed polarizer

Next, the pump-probe arrangement with a nearly-crossed polarizer is considered. In the case, the circuit diagrams remain unchanged and the only difference is the function describing the output detector signal $V_{\text{det}}$ (Table 1); in the electronic-equivalent circuit of the pump-probe arrangement with the nearly-crossed polarizer the detector signal is given by $V_{\text{det}}=V_{\text{probe}}V_{\text{atoms}}$. In the passive mode, where an external generator is used to drive the RLC circuit, the detector signal takes the form $V_{\text{det}}=(\sin\Omega_m t+1)^2=3/2+2\sin\Omega_m t-1/2\cos2\Omega_m t$. It proves an existence of the second harmonic of the modulation frequency $\Omega_m$ in the signal. In the self-oscillating mode, the harmonic ought to be strongly suppressed as the system promotes the RLC-circuit resonance.

Figure 8 shows the results of the numerical simulations. Similarly as before, the system starts up from noise and oscillates at the RLC-circuit resonance [Fig. 8(a)]. The only noticeable departure from the former case is the starting time that is roughly 2 times longer than in the previous arrangement. The simulations also demonstrate that a change in the resonance frequency leads to an instant modification of the oscillation frequency ($t_0$=20 ms) [inset to Fig. 8(a)]. The change is also accompanied by slight decrease of the oscillation amplitude; the maximum oscillations are obtained after a few ms.

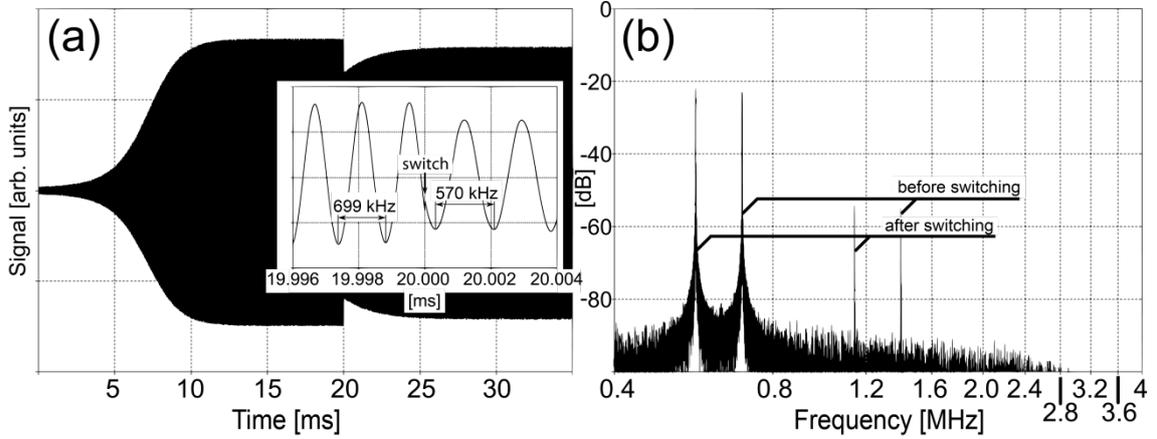

**Figure 8**. (a) Time-dependent signal recorded in the pump-probe arrangement with the nearly-crossed polarizer ($A = \theta = 1$). After an initial starting-up time of about 20 ms the oscillation reaches its maximum. The system instantly responds to the change of the RLC circuit resonance frequency at $t_0$=20 ms (inset). (b) Fourier transform of the signal with clearly visible second harmonics of the resonance frequencies that are absent in the former case. The simulations are performed for the same set of parameters as in Fig. 6.

Figure 8(b) presents the FFT spectrum of the signal shown in Fig. 8(a). The strong two peaks correspond to the oscillations before and after attaching switching capacitor C'. Moreover, the second harmonics of the modulation frequency $\Omega_m$ are observed in the spectrum.

Their amplitudes are orders of magnitude smaller than those of the first harmonics, which is due to spectral filtering of the modulation signal by the RLC circuit.

### 3.3.3 Single-beam arrangement with balanced polarimeter

The same circuits as in the former two cases are also used to simulate the AMOR magnetometer in the single-beam arrangement with the balanced polarimeter. In the arrangement, the light modulation contribute to the detector signal $V_{det}=V_{probe}V_{atoms}$, which, for the passive mode, takes the form $V_{det}=1/2(1+\sin\Omega_m t)\sin\Omega_m t=1/2\sin\Omega_m t-1/4\cos2\Omega_m t$. Operation of the system in self-oscillating mode strongly modifies this relation. In particular, the signal $V_{det}$ contains a significant contribution from $V_{probe}$. Since the modulation signal $V_{probe}$ bypasses the RLC circuit it is not filtered and propagates through the system.

The results of the simulations are shown in Fig. 9. The time-dependent signal significantly deviates from a sine function. This distortion is caused by the application of a multi-harmonic signal to drive the RLC circuit; the complex structure of the modulation propagates through the system and eventually a dynamic equilibrium is obtained after many cycles. The complex character of the signal is best visible on the FFT spectrum [Fig. 9(b)]. As presented, many frequencies are observed before and after attaching the additional capacitor C', which is a result of lack of full spectral filtering of $V_{det}$.

Regardless of the complexity of the simulated signal, the oscillations built up from white noise and the self oscillations are achieved after ~3 ms [Fig. 9(a)]. The instant response of the system to the change of the RLC resonance frequency is also demonstrated [inset to Fig. 9(a)]. It proves the ability of the AMOR magnetometer to track magnetic field changes with no delay even in such a simplified arrangement.

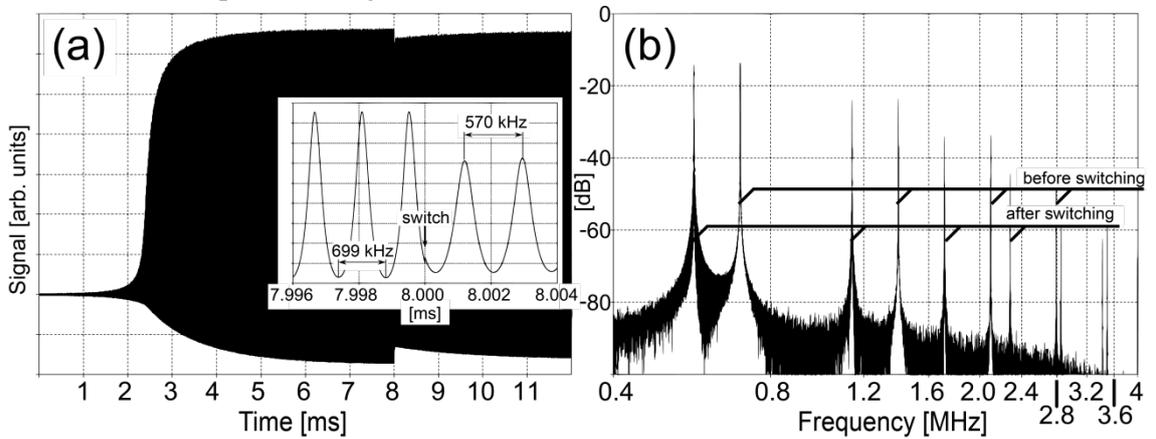

**Figure 9.** (a) Time-dependent signal recorded in the electronic circuit corresponding to the single-beam arrangement with the balanced polarimeter. The signal strongly deviates from the sine function, which does not prevent the system from entering self oscillations. The instant response to the change of the RLC-circuit resonance frequency is shown in the inset. (b) FFT spectrum of the signal with a number of harmonics arising due to complexity of the signal used for driving the oscillations and detecting the response of the system to the modulation.

### 3.3.4 Single-beam arrangement with nearly-crossed polarizer

Due to experimental simplicity the arrangement exploiting a single light beam and a nearly-crossed polarizer is appealing for magnetic-field measurements. Unfortunately, in this configuration the complexity of the signal is the highest among all four arrangements considered in this paper. In the passive mode the signal takes a form $V_{det}=(1+\sin\Omega_m t)^{2}(1+\sin\Omega_m t)/2=1/8(6+7\sin\Omega_m t-2\cos2\Omega_m t-\sin3\Omega_m t)$, showing a number of harmonics present in the signal. Similarly as before, the signal become significantly more complex in the self-oscillating mode, where the modulation propagates through the system.

The diagram of the simulated circuit is presented in Fig. 10. As shown, it differs significantly from the circuits used for the other simulations. For instance, the voltage generator $B_{gen}$ providing voltage $V_{start}$, the control switches $S_2$ and $S_3$, and the voltage sources $V_2$ and $V_3$ are used in the system. Such elements are required, since initial simulations reveal impossibility of building up oscillations from noise; the starting generator is used for a certain period of time at the beginning of the simulations to make system to oscillate. The frequency of the generator needs to be carefully adjusted prior the simulations to fulfill the resonance condition of the RLC circuit to start up the system. It is worth noting that in the simulations the system is most efficiently excited with the generator frequency (699.99 kHz) slightly below the RLC-circuit resonance frequency (700 kHz). The difference is an artifact of simulations rather than real effect as described in Sec. 3.2. After achieving given amplitude of the oscillations, the generator is disconnected by the switch $S_2$ and the switch $S_3$ closes a feedback loop in order to achieve self-oscillations.

Additional modification of the electronic system modeling the magnetometer in this arrangement is introduction of the high-pass filter $R_f$-$C_f$. It is used to prevent propagation of the DC voltage through, in particular entering the feedback loop. The function modeling agc is also modified to ensure continuous oscillations and attention is paid to prevent the detector signal $V_{det}$ from going negative.

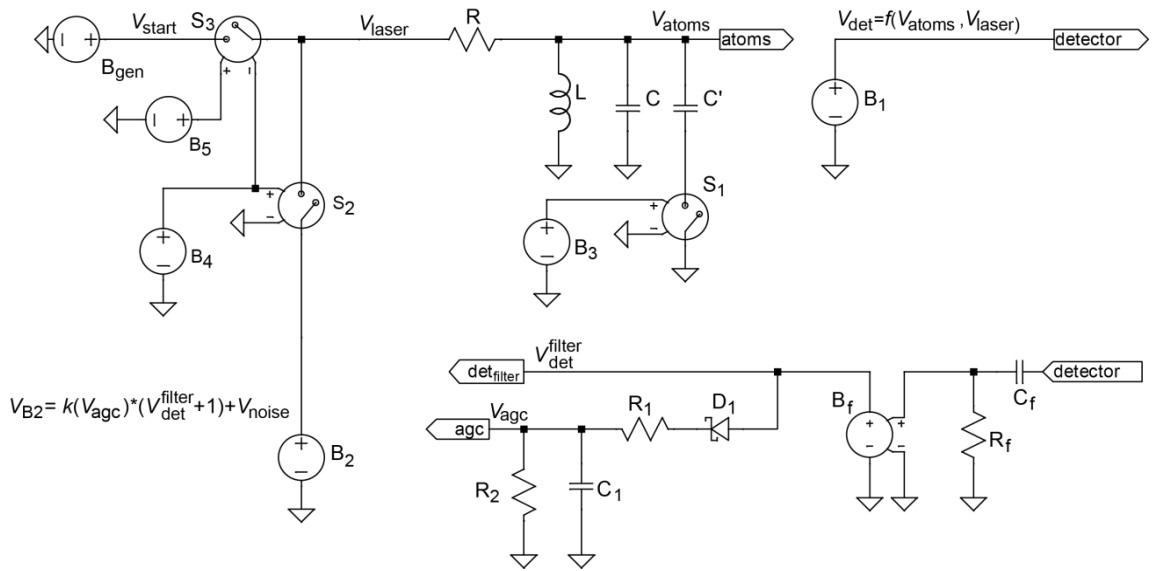

**Figure 10**. Electronic circuits corresponding to the single-beam AMOR magnetometer with the nearly-crossed polarizer. The new element of the system, the high-pass filter that prevents from propagation of the DC signal through a feedback system is shown in bottom right-hand corner of the figure. The agc system is applied to control the amplitude of the modulated signal. For simplicity the voltage $V_{DC}$ protecting from signal to be negative was incorporated in the source $B_2$.

F**igure 11**. (a) Time-dependent signal recorded in a system corresponding to the single-beam arrangement with the nearly-crossed polarizer. After an initial part when the oscillations are excited by the external generator, the system is switched into self-oscillation mode ($t$=4 ms) and high-amplitude self oscillations are obtained. The system oscillates until the resonance frequency of the RLC was changed ($t_0$=6 ms) and oscillations at different frequency with slightly lower amplitude are observed (inset). (b) FFT spectrum of the signal with a number of frequencies appearing in the signal due to application of single light beam and nearly-crossed polarizer.

The simulation results are shown in Fig. 11. For the first 4 ms the system is driven by the generator $B_{gen}$. Strong oscillations right at the start of the simulations are caused exclusively by the modulation of light recorded by the detector. Their following increase is related to the response of the system to the stimulus, i.e., rotation of the polarization. After 4 ms, the feedback loop is closed and the system keeps oscillating. The amplitude of the oscillations slightly varied after the loop is closed but it finally stabilizes using the agc circuit. After another 2 ms ($t$=6 ms) the capacitor $C_b$ is connected to the RLC system simulating the change in the magnetic field in a real magnetometer. The system immediately responds to the change, which demonstrates the potential of the magnetometer in the arrangement for tracking a magnetic field. Amplitude of the observed oscillations depends somewhat on the frequency which is a consequence of the simple function implementing agc. The FFT spectrum presented in Fig. 11(b) reveals a complex structure of the recorded signal.

## 4. Experimental results

In order to experimentally verify the predictions of the simulations and confront them with real-magnetometer readouts, the AMOR magnetometer with a single light beam and balanced polarimeter is set up in the self-oscillation mode. The setup is described in more details in Ref. [6], here we just briefly review its most important features. It consists of a diode laser emitting light tuned and stabilized to the F=2→F'=1 transition in $^{87}$Rb (795 nm), acousto-optical modulator (AOM) applied for modulation of light intensity, balanced polarimeter consisting of a crystal polarizer and two photodiodes monitoring two channels of the polarizer, and the paraffin-coated glass cell filled with rubidium vapor enclosed inside a 3-layer mu-metal shield with precisely controlled magnetic field inside. The difference signal of two polarimeter photodiodes is amplified, phase shifted, and feed into the modulation input of AOM.

The measured signals are shown in Fig. 12. For the sake of convenience the magnetic field is scanned within a range from 1.5 to 2.0 µT which corresponds to the AMOR signal of about 25 kHz. As shown in Fig. 12(a), the signal is periodic with clearly visible second harmonic component of the fundamental frequency. The signal slightly deviates from the one shown in Sec. 3.3.3 but it was experimentally verified that the difference originates from a phase shift present in our system; an appropriate adjustment of the phase delay in our feedback loop resulted in a good agreement between simulations and experimental results.

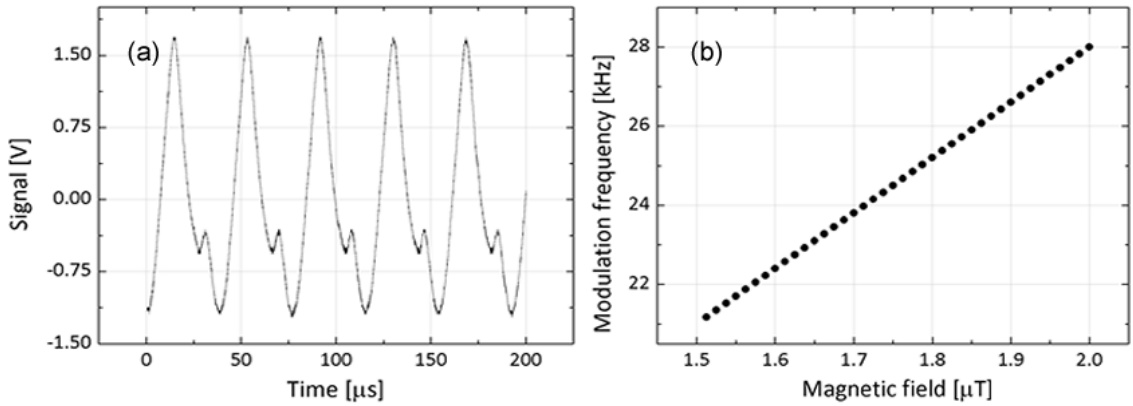

Figure 12. (a) Time-dependent signal recorded in the AMOR magnetometer in the single-beam arrangement with the balanced polarimeter. The signal is a harmonic oscillation with clearly visible two components of the modulation. (b) The magnetic-field tracking signal recorded in the self-oscillating regime.

Figure 12(b) presents the magnetometer tracking signal. As shown, the modulation frequency strictly followed the magnetic field without any external control. It proves that the magnetometer is operated in self-oscillating mode and predictions of our electronic model are valid.

## 5. Conclusions

We designed the electronic model of the optical magnetometer based on amplitude-modulated nonlinear magneto-optical rotation. By using the RLC circuit we model the physical phenomenon underlying the operation of the device and the detection scheme in our experimental apparatus. Four different arrangements of the magnetometer were considered. As demonstrated, in all four arrangements the system may be introduced into the self-oscillating mode and automatic tracking of the resonance frequency of the RLC system and hence a magnetic field may be obtained. We set up the magnetometer in the single-beam balanced-polarimeter configuration and demonstrated the possibility of tracking a magnetic field. Operation of the magnetometer in the self-oscillating regime opens a possibility for its practical application, in particular, for magnetocardiography. Such application is currently under development.

**Acknowledgments**

**Appendix**

The intensity of linearly-polarized light transmitted through a polarizer is given by the Malus law
$$I = I_0\cos^2\alpha, \qquad (2)$$
where $\alpha$ is the angle between polarizer axis and light polarization and $I_0$ corresponds to the incident intensity of light. For a photodiode placed after nearly-crossed polarizer the photocurrent/signal is given by
$$S = I_0\cos^2(\varphi + 90° + \Theta) = I_0\sin^2(\varphi + \Theta) \approx I_0(\varphi + \Theta)^2, \qquad (3)$$
where $\varphi$ is the angle of magneto-optical rotation and $\theta$ is the small angle of rotating the polarizer with respect to 90°. We assume that for small angles $\sin(\varphi+\Theta)\approx(\varphi+\Theta)$.

It can be shown (see, for example, Ref. [13]) that the time-dependent polarization rotation $\varphi(t)$ after propagation through a medium with optical anisotropic that subject a magnetic field is given by
$$\varphi(t) = A\sin 2\Omega_\text{L} t = A\sin\Omega_\text{m} t, \qquad (4)$$
where the resonance condition (1) is used for last equality. Substituting Eq. (4) into Eq. (3) one gets
$$S = I_0(t)(A\sin\Omega_\text{m} t + \Theta)^2. \qquad (5)$$
Depending on a time dependence of the incident light intensity $I_0(t)$, which is constant for pump-probe arrangements and modulated as $(1+\sin\Omega_\text{m}t)/2$ for single-beam arrangement, the second and fourth row formulas in Table 1 are obtained.

In case of the balanced polarimeter the output signal is a difference of photocurrents of two photodiodes monitoring the channels of crystal polarizer, which optical axis is rotated by 45° with incident polarization
$$S = I_1 - I_2 = I_0[\cos^2(45° + \varphi) - \cos^2(45° - \varphi)] = I_0/2[\sin 2\varphi - \sin(-2\varphi)] \approx 2I_0\varphi, \qquad (5)$$
where trigonometric relation $\cos^2\alpha=(1-\cos 2\alpha)/2$ is used. Substituting Eq. (4) into Eq. (5) one finds
$$S = 2I_0(t)\sin\Omega_\text{m} t. \qquad (6)$$
Introducing different time dependence of the incident light, the third and fourth formulas in third column of Table 1 are obtained.

Finally, it is noteworthy that in three out of four cases considered in this paper, the signal S contains terms at higher power of $\sin\Omega_\text{m}t$. Based on trigonometric transformation one can shown that such terms manifest as harmonics of $\Omega_\text{m}$ in the signal.